\documentclass[aps,twocolumn,pre,notitlepage,10pt]{revtex4-1}

\usepackage{graphicx}
\usepackage{mathtools}
\usepackage{amssymb}
\usepackage{MnSymbol}
\usepackage{pdfsync}
\usepackage{color}

\newcommand{\grad}{\boldsymbol{\nabla}}

\newcommand{\bF}{{\mathbf F}}
\newcommand{\bff}{{\mathbf f}} 
\newcommand{\bN}{{\mathbf N}} 
\newcommand{\bn}{{\mathbf n}}

\newcommand{\br}{{\mathbf r}} 
\newcommand{\bT}{{\mathbf T}} 
\newcommand{\bt}{{\mathbf t}}

\begin{document}

\author{Ciro Semprebon$^{1}$, Mario Scheel$^{1,2}$, Stephan
  Herminghaus$^{1}$, Ralf Seemann$^{1,3}$, and Martin Brinkmann$^{1,3}$}

\affiliation{(1) Max-Planck-Institute for Dynamics and
  Self-Organization, Am Fassberg 7, D-37077 G\"{o}ttingen, Germany}
\affiliation{(2) Synchrotron Soleil, L’Orme des Merisiers,
  Saint-Aubin, F-91192 Gif-sur-Yvette, France}
\affiliation{(3) Experimental Physics, Saarland University, D-66123
  Saarbr\"ucken, Germany}
\email[]{martin.brinkmann@physik.uni-saarland.de}

\title{Liquid morphologies and capillary forces between three
  spherical beads}

\date{\today}

\begin{abstract}
  Equilibrium shapes of coalesced pendular bridges in a static
  assembly of spherical beads are computed by numerical minimization
  of the interfacial energy. Our present study focuses on generic bead
  configurations involving three beads, one of which is in contact to
  the two others while there is a gap of variable size between the
  latter. In agreement with previous experimental studies, we find
  interfacial `trimer' morphologies consisting of three coalesced
  pendular bridges, and `dimers' of two coalesced bridges. In a
  certain range of the gap opening we observe a bistability between
  the dimer and trimer morphology during shrinking and growth. The
  magnitude of the corresponding capillary forces in presence of a
  trimer or dimer depends, besides the gap opening only on the volume
  or Laplace pressure of liquid. For a given Laplace pressure, the
  capillary forces in presence of a trimer are slightly larger than
  the force of a single bridges at the same gap opening, which could
  explain the shallow maximum and plateau of the capillary cohesion of
  a wetting liquid for saturations in the funicular regime.
\end{abstract}

\maketitle

\section{Introduction}
\label{sec:introduction}

Capillary forces and the mechanics of wet granular materials have
remained an active field of research since the seminal works of Haines
\cite{Haines1925} and Fisher \cite{Fisher1926} on soil mechanics. Most
models consider capillary forces in the presence of cylindrically
symmetric pendular bridges \cite{Gillespie1967, Hotta1974, Orr1975,
  Lian1993, Willett2000, Dorrmann2014}, and are thus restricted to the
describe the mechanics at low liquid saturations
\cite{Richefeu2006,Mani2012}. Recent developments in three dimensional
imaging techniques such as fast confocal microscopy \cite{Butt2009}
and X-ray tomography \cite{Scheel2008a} made it possible to
investigate the shape of liquid clusters in the funicular regime with
a resolution well below the scale of a single grain. Despite these
advances, only a few attempts have been made to quantify capillary
forces in the funicular regime
\cite{Mitarai2006,Scheel2008a,Herminghaus2013}. It is evident that the
magnitude of the cohesive forces at different liquid saturations is
intimately linked to the morphology of the interstitial fluids on the
scale of single grains \cite{Scheel2008a,Melnikov2015}. A better
understanding of the liquid structures emerging in the funicular
regime and the capillary cohesion caused by them may help to predict
landslides or avalanches \cite{Halsey1998}, and to complement existing
models for technological applications in wet aggregation or particle
coating \cite{Turton2008}. Besides the mechanics of wet granulates,
modeling the cluster morphology of partially wetting liquids will also
have repercussions on the theory of fluid transport in wet granular
beds
\cite{Herminghaus2005,Scheel2008b,Mani2012,Mani2015,Melnikov2015}.

In this article we present the results of a systematic study of
fundamental liquid morphologies and corresponding capillary forces in
arrangements of spherical beads beyond the regime of pendular
bridges. These clusters appear at liquid saturations corresponding to
the transition form the pendular to funicular regime. Considering
perfectly wetting liquids and a random close packing of spherical
beads, the pendular bridge regime is limited to a range of liquid
content $W<W^\ast\approx 2.5\cdot 10^{-2}$, here expressed with
respect to the total sample volume \cite{Harris1964,Scheel2008a}. At
slightly higher liquid contents $W>W^\ast$, the lateral extension of a
pendular bridges on the bead surface does not anymore permit the
formation of isolated bridges. Consequently, a certain fraction of the
pendular bridges coalesces and transforms into a funicular structure,
i.e.~into liquid clusters that are simultaneously in contact to at
least three beads.

\begin{figure}
  \centering
  \includegraphics[width=0.8\columnwidth]{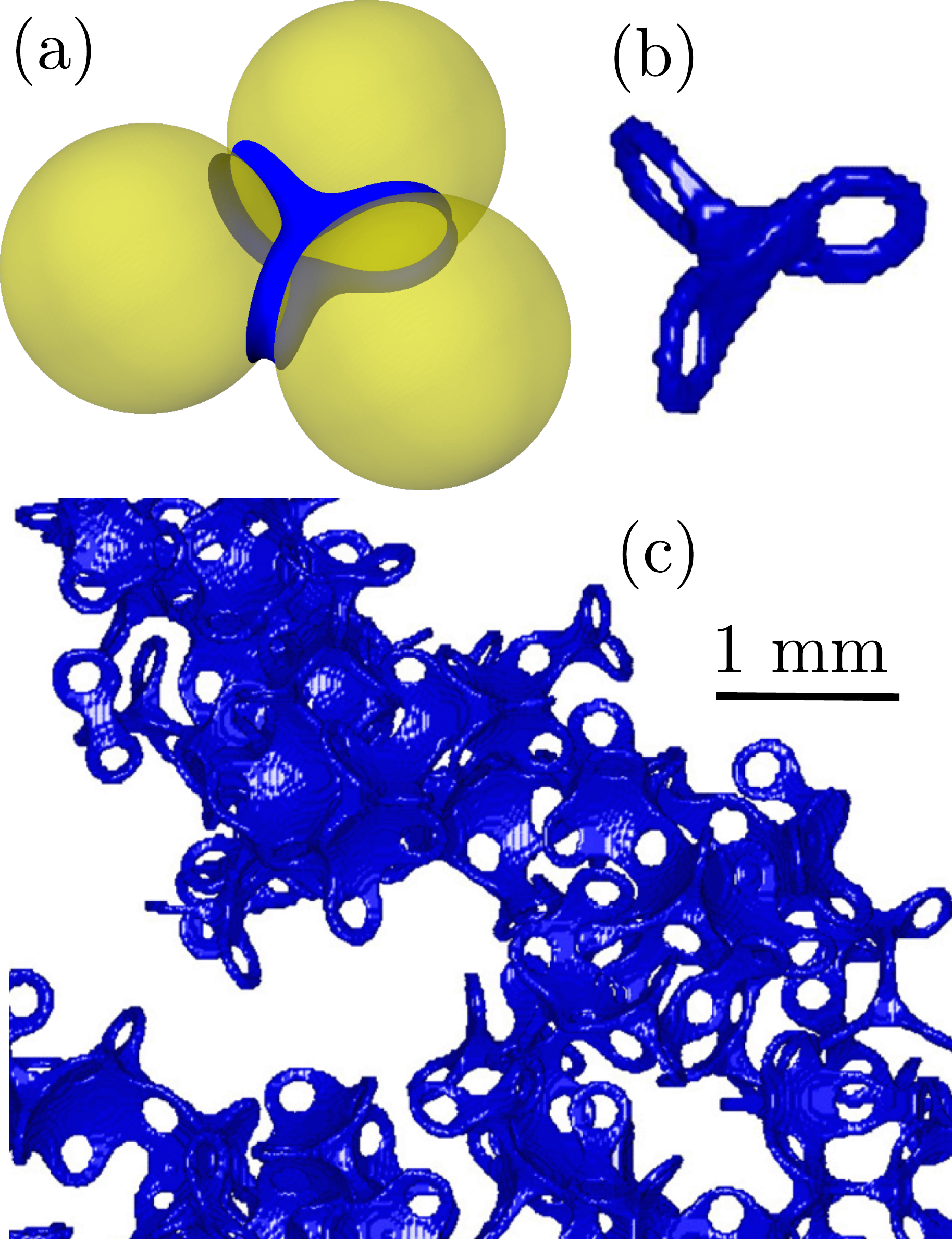}
  \caption{(Color online) (a): Trimer morphology computed in our
    numerical energy minimizations. (b) and (c): Renderings of
    segmented X-ray mircotomography images of a wetting liquid between
    spherical glass beads and a large liquid cluster in a disordered
    packing from x-ray tomography \cite{Scheel2008a}. Bead radius
    $R=(355\pm 35)\mu$m, wetting liquid aqueous ZnI$_2$ solution in
    (b) and (c).}
  \label{fig:cluster_tomos}
\end{figure}

The most characteristic cluster morphology found in granular bed of
highly wettable spherical beads is a `trimer' of pendular bridges
\cite{Scheel2008a}.  As can be seen in the three dimensional rendering
of x-ray tomography data in Fig.~\ref{fig:cluster_tomos}~(a),
\cite{Scheel2008a} three pendular bridges have coalesced around a
triangular opening formed by three adjacent beads. In what follows we
refer to the center of the opening as a `throat'. Throats are found in
large numbers in a disordered assembly of spherical beads
\cite{Finney1970,Scheel2008a}, and it is not surprising that filled
throats connecting three adjacent bridges represent the most generic
liquid structure beyond isolated pendular bridges. A close inspection
of the liquid morphology shown in Fig.~\ref{fig:cluster_tomos}~(c)
reveals that this large liquid cluster indeed consists mostly of these
trimer units: Any of the three pendular bridges belonging to a certain
trimer unit can be part of a least one more trimer unit. The filled
throats in a granular assembly can thus form large interconnected
chains of pendular bridge `polymers'. Owing to its outstanding
importance we will focus our present numerical study on a single
trimer.

Capillary forces of pendular bridges between two spherical beads in
contact display only small variations with the liquid volume
\cite{Fisher1926,Halsey1998,Herminghaus2013}. This observation readily
explains the insensitivity of the mechanical cohesion of a wet
granular assembly with respect to the liquid content $W$ in the
pendular regime $W\lesssim W^\ast$ \cite{Halsey1998}. Quite
surprisingly, the magnitude of capillary cohesion evolves smoothly
while the liquid content is increased from the pendular regime
$W\lesssim W^\ast$ into the funicular regime $W\gtrsim W^\ast$
\cite{Scheel2008a}. Measurements of capillary cohesion employing the
fluidization threshold or the tensile strength
\cite{Richefeu2006,Scheel2008a} reveal a plateau in the range of $W$
from approximately $0.03$ reaching to values larger than $\approx
0.1$. At higher liquid contents, capillary cohesion gradually weakens
with increasing $W$ and vanishes at full saturation of the pore space
$(W \approx 0.4)$. In the latter case, every bead in the granular bed
is completely immersed in liquid and, thus, there are no fluid
interfaces that can give rise to capillary cohesion.

An experimental quantification of capillary forces in small ensembles
of beads is difficult and {\em in situ} measurements are hardly
feasible with the currently available experimental techniques. To
reach a quantitative picture of capillary cohesion between three or
more beads, we employ numerical minimizations of the interfacial free
energy and obtain the equilibrium morphology of the liquid interfaces
along with the magnitude and direction of capillary forces. The
respective capillary forces as a function of the surface--to--surface
separations can be then employed to formulate a model for the
mechanics of wet beads, provided the complete set of rules to
transform, merge, or split liquid cluster in response to slow changes
of the local bead configuration is known \cite{Melnikov2015}.

A statistical analysis of X-tomography images revealed that local
triangular configurations with three detectable finite separations
between the bead surfaces are extremely rare in random packings of
monodisperse beads \cite{Scheel2008a,Scheel2008b}. The majority of
these local `throats' exhibit only one detectable gap between the
surfaces of the beads. The beads of the remaining two pairs are in
mechanical contact. The other extreme case where all three bead pairs
are in contact is equally rare. Significant differences in the
statistics of surface-to-surface separations of adjacent beads between
dry and wet assemblies could be detected in the X-ray tomography
\cite{Scheel2008a}. In view of the three dimensional parameter space
that accounts for the most general configurations of three beads, we
will restrict our discussion in this work to the most relevant case,
i.e.~three bead configurations with two mechanical contact and one
finite gap opening. To account for the range of experimentally
relevant situations we report the results for both the volume
controlled case and the pressure control case. Local bead
configurations where all three pairs are in contact are covered as a
special case. In a forthcoming work, we will present the results on
the general case with two and three finite gap openings, respectively.

This article is organized as follows: In
Sec.~\ref{sec:numerical_model} we introduce the physical model and
give details about the numerical method to compute equilibrium
configurations of the liquid interface. In
Sec.~\ref{sec:numerical_model}, we summarize the fundamentals of
theory of capillarity and methods to compute capillary forces between
solid bodies in numerical energy minimizations. The appearance of
trimers and dimers of pendular bridges in local bead configurations
with two contacts are described in
Sec.~\ref{subsec:liquid_morphologies} and compared to wetting
experiments in Sec.~\ref{subsec:experiments}. Numerically computed
capillary forces between beads in the presence of trimers and dimers
are presented and discussed in
Sec.~\ref{subsec:capillary_forces}. Finally, we conclude with a
summary and outlook in Sec.~\ref{sec:outlook}.

\section{Physical model and numerical implementation}
\label{sec:numerical_model}

Before we address the theory of capillary forces, we will first give a
short outline of the framework of capillarity. To this end, we will
consider the most general case of $N$ spherical beads in contact to
two fluid phases. The position of the $N$ bead centers are given by
vectors ${\mathbf r}_1, {\mathbf r}_2, ..., {\mathbf r}_N$ with
Cartesian coordinates ${\mathbf r}_i=(x_i,y_i,z_i)$. For later
convenience we denote vectors and tensors by boldface symbols. The
fluid phases in contact to the beads may represent two immiscible
liquids or a liquid in coexistence with a vapor phase. As we assume
the beads to be fixed in space, or allowing only adiabatically slow
variations of their positions, any flow inside the fluid phases has
ceased once a mechanical equilibrium of the bulk phases and their
interfaces has been reached. In this state, both fluids are at rest
relative to the surface of the beads which implies the absence of
viscous stresses.

\subsection{Interfacial energies}
\label{subsec:interfacial_energies}

The shape of the liquid-vapor interface in mechanical equilibrium and
the corresponding capillary forces and torques acting on the beads can
be obtained from thermodynamic considerations of the interfacial free
energy ${\cal E}$. For given positions
$\br\equiv(\br_1,\br_2,....,\br_N)$ of the bead centers, the
functional
\begin{equation}
  {\mathcal E}\{{\Sigma^{\rm lv}}\} =\gamma^{\rm lv}\,A^{\rm lv}
  +\left(\gamma^{\rm ls} -\gamma^{\rm vs}\right) \sum_{i=1}^N\,A^{\rm
    ls}_i
  \label{eq:energy_functional_E}
\end{equation} 
assigns a free energy to every configuration $\Sigma^{\rm lv}$ of the
liquid-vapor interface. The first term on the RHS of
eqn.~(\ref{eq:energy_functional_E}) represents the free energy of the
liquid-vapor interface with area $A^{\rm lv}$ and interfacial tension
$\gamma^{\rm lv}$. The second term in
eqn.~(\ref{eq:energy_functional_E}) accounts for the free energy
related to the surface $\Sigma^{\rm ls}_i$ of bead $i$ in contact to
the liquid with area $A^{\rm ls}_i\equiv |\Sigma^{\rm ls}_i|$ and the
liquid-solid and vapor-solid surface tension $\gamma_{\rm ls}$ and
$\gamma_{\rm vs}$, respectively.

Throughout this work, we will neglect the influence of gravity and
buoyancy, as their effects are typically small in liquid clusters with
extensions on the length scale of the grains. The capillary length for
an air water interface is in the range of a few
millimeters \footnote{To account for gravity or buoyancy one needs to
  add a term $\Delta\rho\,g\,z_{\rm cm}$, where $\Delta \rho\equiv
  \rho^{\rm l}-\rho^{\rm v}$ is the density difference between the
  liquid and the vapor phase, $g$ the acceleration of gravity, and
  $z_{\rm cm}$ the vertical center of mass position of the liquid
  body.}. Expressed in terms of the Bond number $\mathsf{Bo} \equiv
\Delta \rho \, g R^2 /\gamma^{\rm lv}$ with a typical vertical
dimension of the interface in the range of the bead radius, we have
$\mathsf{Bo}\ll 1$. Furthermore, we assume that the typical distances
between the liquid interface and solid walls are large such that
contributions of the disjoining pressure to the normal stress at the
interface can be safely excluded.

Under these conditions, any mechanically stable interface
configuration $\Sigma^{\rm lv}$ in contact to the beads with fixed
positions $\bar\br$, is a local minimum of the free energy
eqn.~(\ref{eq:energy_functional_E}). Dealing with non-volatile
liquids, we have to consider minima of the functional
(\ref{eq:energy_functional_E}) under the subsidiary constraint of a
fixed volume $V$ of the liquid body. We will refer to this situations
as the `volume controlled case'.

In some situations it is justified to assume that the liquid in
contact to the beads can be exchanged with a reservoir that fixes the
pressure difference $P\equiv P^{\rm l}-P^{\rm v}$ between the liquid
(l) and vapor (v) phase. If this is the case we have to find local
minima of the Grand interfacial free energy
\begin{equation}
  {\cal G}={\cal E}-P\,V~,
  \label{eq:energy_functional_G}
\end{equation}
The grand interfacial free energy takes into account the work received
from or done at the volume reservoir, respectively. We will refer to
this case as the `pressure controlled case'.

Any extremum of the free energy ${\cal E}$ under the constraint of a
fixed volume $V$ necessarily satisfies two conditions. The first
condition is expressed by the Young-Laplace equation:
\begin{equation}
  P=2H\gamma^{\rm lv}~,
  \label{eq:Laplace_pressure}
\end{equation}
that holds in every point of the interface $\Sigma_{lv}$. The mean
curvature $H$ is the sum of the two principal curvatures, or any pair
of normal curvatures of the interface into orthogonal direction, see
for instance Ref.~\cite{doCarmo1976}. In the absence of gravity, the
Laplace pressure $P$ is independent on the position which implies that
$\Sigma^{\rm lv}$ is a surface of constant mean curvature.

The second necessary condition of a local free energy minimum is due
to Young, Dupr\'e, and Laplace and expresses the mechanical
equilibrium at every point the three phase contact line:
\begin{equation}
  \gamma^{\rm lv}\cos\theta_0=\gamma^{\rm sv}-\gamma^{\rm sl}~.
  \label{eq:Young_Dupre}
\end{equation}
The equilibrium contact angle $\theta_0$, also termed Young's or
material contact angle, is determined solely by interfacial free
energies. For comparison to `real' experimental systems, we have to
consider static advancing and static receding contact angles instead
of Young's angle $\theta_0$ to account for the dissipation of work
during and advancing or receding motion of the three phase contact
line on a heterogeneous surface.

Owing to inherent non-linearity of the energy functional
eqn.~(\ref{eq:energy_functional_E}), the interfacial free energy
${\cal E}$ may exhibit more than a one local extremum for a given
liquid volume $V$ and given positions $\bar\br$ of the beads. The
equilibrium shapes can be distinguished by suitable order parameter
describing the interfacial shape. The free energy ${\cal E}$ of these
equilibrium shapes is a function $E(\bar\br,V)$ in a certain range of
bead coordinates $\bar\br$ and liquid volume $V$.

In the many instances one finds multiple local minima for given
positions $\bar\br$, volume $V$, and material contact angle
$\theta_0$. In this case, the function $E(\bar\br,V)$ is multi-valued
and forms a number of branches (or `leafs'). Analogous statements hold
for the Grand free energy ${\cal G}$ and the corresponding energy
landscape $G(\bar\br, P)$, where the Laplace pressure $P$ represents
the accessible control parameter, instead of the liquid volume $V$.

Any interfacial configuration that satisfies
eqns.~(\ref{eq:Laplace_pressure}) and (\ref{eq:Young_Dupre}) is a
extremum of the interfacial energy ${\cal G}$ for the given Laplace
pressure $P$ in eqn.~(\ref{eq:Laplace_pressure}). From this
observation we can conclude that an extremum of the free energy
eqn.~(\ref{eq:energy_functional_E}) for a certain volume $V$ is also
an extremum of the Grand free energy
eqn.~(\ref{eq:energy_functional_G}) for a certain pressure $P$ and
vice versa. This implies that the set of interfacial equilibria in the
volume controlled case and in the pressure controlled case are
identical. The mechanical stability of these extrema, however, may
differ between the volume and the pressure controlled cases. Apart
from a constant, the free energies ${\cal G}$ and ${\cal E}$ are
identical for all interfacial configurations enclosing the same volume
$V$. Hence, any local minimum of ${\cal G}$ will be also a local
minimum of ${\cal E}$, where the Laplace pressure $P$ in ${\cal G}$
can be regarded as a Lagrange multiplier to enforce the constraint of
a constant liquid volume $V$. Mechanical stability in the pressure
controlled case always implies mechanical stability in the volume
controlled case. The converse statement, however, is not necessarily
true: the liquid may exhibit more locally stable states in the volume
controlled case as compared to the pressure controlled case.

 At specific values of the relevant control parameters liquid volume
 $V$ or Laplace pressure $P$ and bead configuration $\bar{\bf r}$,
 termed `bifurcation points', the number of local minima and saddle
 points of the energy functionals $\cal E$ and $\cal G$
 changes. Bifurcation points can be classified according to universal
 aspects of the underlying energy landscape, i.e.~the number of
 control parameters and order parameters as well as the symmetries of
 the energy functionals is the subject of catastrophe theory,
 cf.~Refs.~\cite{Thom1989,Arnold2003}.

\begin{figure}
  \includegraphics[width=0.7\columnwidth]{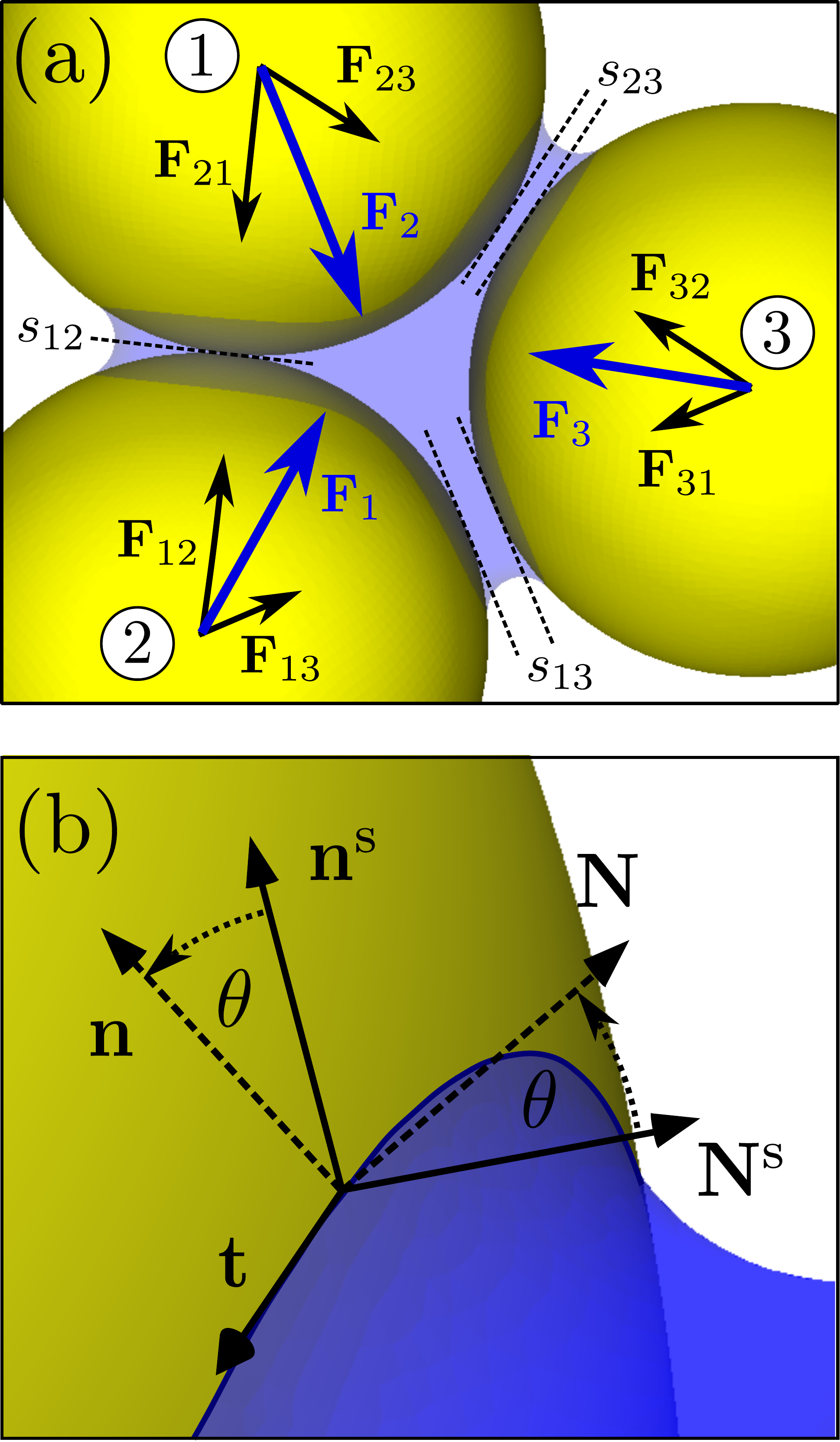}
  \caption{(Color online) (a): Geometry of three spherical beads with
    arbitrary gap openings $s_{ij}$, $i \neq j\in\{1,2,3\}$. The total
    capillary force $\bF_i$ acting on sphere $i$ (blue arrows) is
    decomposed into force pairs acting along the line joining the bead
    centers (black arrows). (b):Local orthonormal vector bases in a
    point $\br$ of the contact line $\Gamma_i$ of a liquid-vapor
    interface $\Sigma^{\rm lv}$ on the surface $\Sigma^{\rm s}_i$ of
    beads $i$. Here, $\bt$ denotes the tangent to $\Gamma_i$, $\bN_s$
    the normal of $\Sigma^{\rm s}_i$, $\bN$ the normal of $\Sigma^{\rm
      lv}$, $\theta$ the contact angle, while $\bn$ and $\bn^{\rm s}$
    denote the outward pointing co-normal vectors of $\Sigma^{\rm lv}$
    and $\Sigma^{\rm ls}$, respectively.}
  \label{fig:force_decomposition_trimer}
\end{figure}

\subsection{Capillary forces}
\label{subsec:capillary_force}

The interfacial free energy $E(\bar\br,V)$ of a branch of liquid
equilibrium states can be employed to compute the capillary force
acting on the beads. Besides a negative sign, the total force $\bF_i$
acting on bead $i$ can be obtained from a gradient
\begin{equation}
  \bF_i=-\grad_{{\br}_i}\,E(\bar\br, V)~,
  \label{eq:total_force_energy_E}
\end{equation}
with respect to the Cartesian coordinates $\br_i$ of the particles
$i\in\{1,2,...,N\}$.

Capillary forces are unique for the equilibrium conformation of the
liquid in contact to the beads, and do not depend on whether the
liquid volume or the pressure is considered as the controlled
parameter. Whenever $\partial_V \tilde P(\bar\br,V)\neq 0$, we can
find the inverse function $V=\tilde V(\bar\br,P)$ for given bead
positions $\bar\br$, where we use the tilde to distinguish the
variable from the respective function. As shown in the appendix
\ref{subsec:ensemble_capillary_forces}, we can derive the identity
\begin{equation}
  \grad_{{\br}_i}G\big(\bar\br, P \big)=
  \grad_{{\br}_i}E\big(\bar\br,V \big)|_{V=\tilde V(\bar\br, P)}~,
  \label{eq:total_force_energy_equiv}
\end{equation}
expressing that the capillary forces in a mechanical equilibrium are
the same in the volume controlled case and the pressure controlled
case, provided we consider corresponding values of the volume $V$ and
the Laplace pressure $P$. Higher derivatives of $E(\bar\br,V)$ and
$G(\bar\br,P)$ with respect to the coordinates $\bar\br$ (evaluated at
a Laplace pressure $P=\tilde P(\bar\br,V)$), however, differ in the
general case.

Once the interfacial energies $\tilde E(\bar\br, V)$ or $\tilde
G(\bar\br, P)$ are known, we can derive the capillary force $\bF_i$
from partial derivatives of with respect to the Cartesian coordinates
$r_{i\alpha}$, $\alpha\in\{x,y,z\}$ for every bead $i$. This approach
is numerically costly because one needs to first scan a high
dimensional function over a large range of parameter. Calculating the
capillary forces directly from the configuration $\Sigma^{\rm lv}$ of
the interface in a local minimum of ${\cal E}$ or ${\cal G}$,
respectively, involves integrations over the three phase contact line
of the interface which is much less costly.

To determine the total capillary force acting on each single bead, let
us first imagine that every bead is enclosed by a control
surface. This virtual control surface shall be infinitesimally close
to the bead surface. By definition, the control surface must not
intersect with any of the adjacent beads. Integration of the normal
component of the stress tensor over the control surface yields the
total capillary force on a bead. The magnitude and direction of this
force is independent on the particular choice of the control surface.

The the total capillary force $\bF_i$ acting on bead $i=1,2,3$ can be
split into a contribution of the isotropic pressure in the liquid and
ambient fluid (index p) and a contribution that stems from the
interfacial tension (index t):
\begin{equation}
  \bF_i=\bF^{\rm p}_i+\bF^{\rm t}_i~.
\end{equation} 
The first contribution $\bF^{\rm p}_i$ is given by the Laplace 
pressure $P\equiv P^{\rm l}-P^{\rm v}$ multiplied by the local surface
normal $\bN^{\rm s}$ of the solid, and integrated over the surface
$\Sigma^{\rm ls}_i$ of the bead $i$ in contact to the liquid:
\begin{equation}
  \bF^{\rm p}= \int_{\Sigma^{\rm ls}_i}{\rm d}A\;P\,\bN^{\rm s}~.
  \label{eq:force_pressure_contribution}
\end{equation}
Note, that this contribution to the capillary force vanishes if the
bead is completely immersed in the liquid provided that $P$ is
constant.

The second contribution $\bF^{\rm t}_i$ arises only in the presence of
a three phase contact line $\Gamma_i$ on the bead $i$, i.e.~if bead
$i$ is partially wet. Locally, the interfacial tension $\gamma^{\rm
  lv}$ of the liquid-vapor interface $\Sigma^{\rm lv}$ pulls into a
direction perpendicular to both the tangent vector $\bt$ of the
contact line $\Gamma_i$, and the local normal vector $\bN$ of
$\Sigma^{\rm lv}$ in a point $\br$ on $\Gamma_i$, which is expressed
as a force per unit length
\begin{equation}
  \bff= \gamma^{\rm lv}(\bN^{\rm s}\sin\theta-\bn^{\rm s}\cos\theta)~.
  \label{eq:line_force}
\end{equation}
The sketch in Fig.~\ref{fig:force_decomposition_trimer}(b) illustrates
the definition of the two local orthonormal vector bases $\{\bt, \bN,
\bn \}$ of the liquid vapor interface $\Sigma^{\rm ls}$ and $\{\bt,
\bN^s, \bn^s \}$ on the surface $\Sigma^{\rm s}$ of the bead, as well
as the local contact angle $\theta$ between $\Sigma^{\rm ls}$ and
$\Sigma^{\rm s}$ in a point $\br \in \Gamma_i$. The surface normal of
the free interface, $\bN$, and the local normal of the bead surface,
$\bN^{\rm s}$, allow us to express the local contact angle as
\begin{equation}
  \cos\theta=\bN\cdot\bN^{\rm s}~.
\end{equation}
The vectorial line force eqn.~(\ref{eq:line_force}) integrated over
the contact line $\Gamma_i$ on bead $i$ yields the total force that
the liquid-vapor interface exerts on bead $i$:
\begin{equation}
  \bF^{\rm t}_i=\int_{\Gamma_i} {\rm d}\ell\;\bff(\br)~.
  \label{eq:force_interfacial_tension_contribution}
\end{equation}

In our numerical energy minimizations, integrals of
eqn.~(\ref{eq:line_force}) over the contact line can be performed as
summations over the corresponding expressions at the edges
representing the contact line. Employing the divergence theorem, one
can re-express the surface integral in
eqn.~(\ref{eq:force_pressure_contribution}) by line a integral over
suitable functions.

In the remainder of this article, we will consider the ideal case of a
homogeneous bead surfaces where the local contact angle is identical
to Young's angle $\theta_0$ as given by eqn.~(\ref{eq:Young_Dupre}).

\begin{figure}
  \includegraphics[width=0.85\columnwidth]{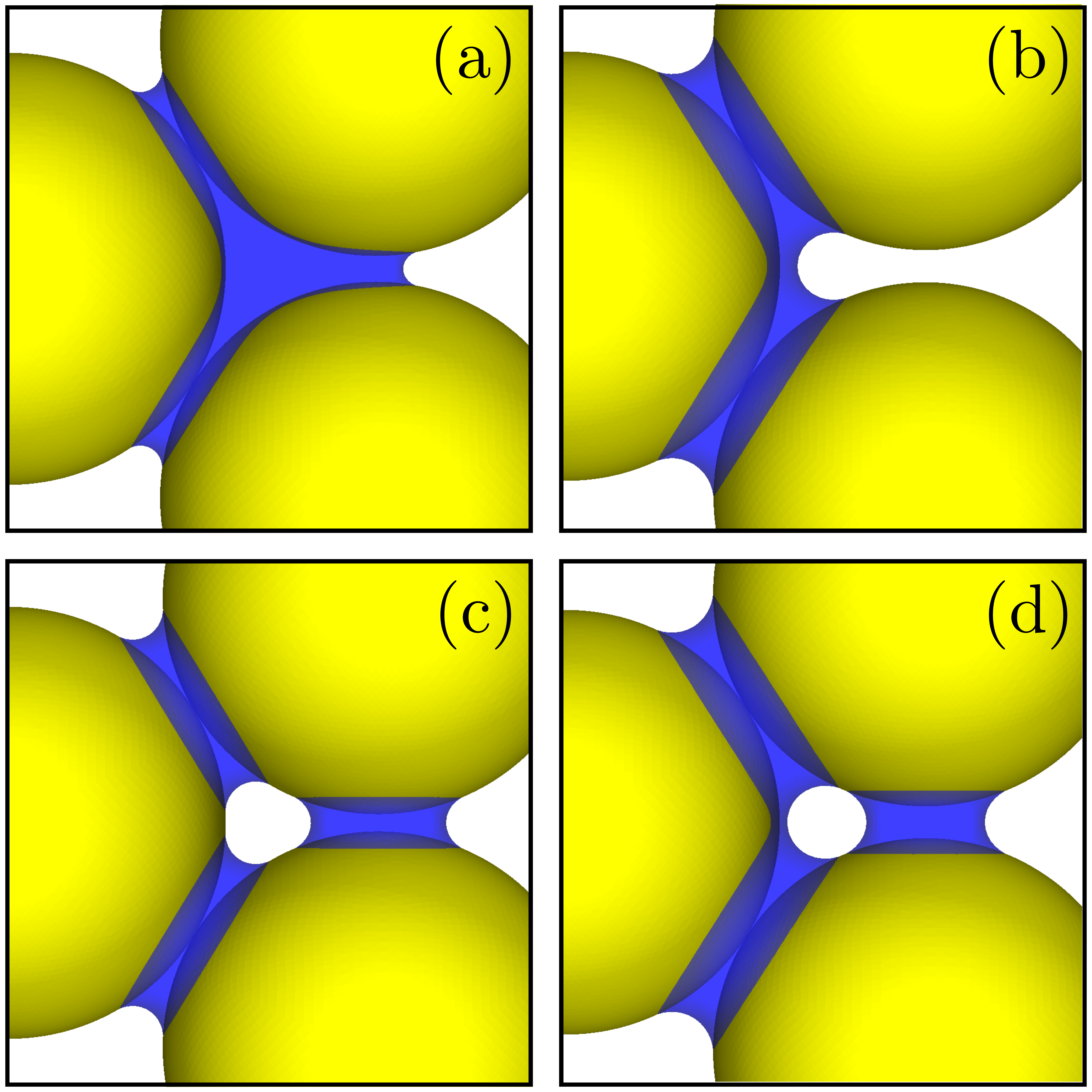}
  \caption{(Color online) Liquid morphologies in contact to three
    beads of equal radius with one non-zero gap opening showing (a) a
    trimer, (b) a dimer, (c) three pendular bridges, and (d) a dimer
    coexisting with a single pendular bridges.}
  \label{fig:morphologies_one_gap}
\end{figure}

\subsection{Numerical implementation}
\label{subsec:numerical_implementation}

Due to a lack of high symmetries like, e.g.~the rotational symmetry,
the computation of the equilibrium shapes of liquid volume wetting
three spheres can be achieved only by means of numerical methods. In
this work we employ a numerical minimization of the interfacial energy
with the freely available software Surface Evolver developed by
K.~Brakke~\cite{Brakke1990}. A fundamental idea behind this method is
that the shape a liquid droplet in contact to a rigid substrate is
entirely fixed by the configuration of the free liquid-vapor
interface.

In the numerical model of the liquid droplet, or droplets in contact
to the beads the shape of the liquid-air interface is approximated by
a mesh of small triangles spanning a set of nodes. The interfacial
energy is a function of the $3N$ coordinates of the nodes, and can be
minimized using a number of build in optimization algorithms
\cite{Brakke1990}. Nodes of the mesh which represent the three phase
contact line have to glide and stay on the spherical surface of one of
the beads. Local geometrical constraint keeps the nodes of the contact
line on the surface. Contributions to the interfacial energy which
stem from the surface of the beads in contact to the liquid are
completely determined by the configuration of the boundary to the
liquid-air interface and are numerically computed from line integrals
of suitably chosen functions over the closed contact line
\cite{Brakke1990}.

During minimization of the interfacial energy, the condition of a
constant liquid volume is imposed through a global integral
constraint. Provided the configuration is in a local minimum of the
interfacial free energy, the Lagrange multiplier corresponding to this
integral constraint that has to be calculated in every minimization
step is identical to the Laplace pressure $P$, i.e.~the pressure
difference across the liquid-air interface. Alternative to the
ensemble of interfacial configurations enclosing a constant volume,
one may consider an ensemble where the Laplace pressure of the liquid
is fixed by a reservoir while the volume is allowed to fluctuate.

A variety of gradient based energy minimizations schemes including
conjugate gradient descent are implemented in the surface evolver.
Furthermore, a complete script language allows us to extract
geometrical quantities from the liquid-air interface, such the
position of special points of the contact line. Long and short edges
of the triangulation are refined, respectively, removed from the mesh
after a number of minimization steps in order to keep the size
distribution of the triangles in a desired range. Subsequent edge
flipping allows the mesh to adapt to large changes of the
configuration.

Preforming numerical computations of interfacial energies, Laplace
pressures, and forces, it is useful to employ dimensionless rescaled
physical quantities. For later convenience, we rescaled any length in
the system by $L_0\equiv R$ and express volumes in units of $V_0
\equiv R^3$. Energies will be rescaled by $E_0\equiv\gamma^{\rm lv}
R^2$ and, consequently, capillary forces and the Laplace pressure by a
force scale $F_0\equiv \gamma^{\rm lv}\,R$ and pressure scale
$P_0\equiv\gamma^{\rm lv}/R$, respectively. For the sake of brevity
and to improve the readability of the text, we will from now on speak
of the non-dimensional rescaled quantities, if not otherwise
mentioned.

\section{Results}
\label{sec:results}

Throughout this work we consider three identical spherical beads, as
illustrated in the sketch of
Fig.~\ref{fig:force_decomposition_trimer}. Assuming a wetting liquid
in our numerical energy minimizations wet set the contact angle to
$\theta_0=5^\circ$. Two of the three bead pairs in the numerical model
are in contact, while the third pair exhibits a finite gap. Without
losing generality, we chose gap openings $s_{12}>0$ and
$s_{13}=s_{23}=0$, i.e.~bead $1$ and $2$ are not in mechanical
contact. Hence, the two remaining control parameter that define the
relative position of the beads are $s_{12}$, and either the liquid
volume $V$ or the Laplace pressure $P$.

\subsection{Liquid morphologies}
\label{subsec:liquid_morphologies}

Figure \ref{fig:morphologies_one_gap} presents four mechanically
stable interfacial morphologies encountered in our numerical energy
minimizations in the volume controlled case. These are the trimer of
three coalesced pendular bridges in panel (a) of
Fig.~\ref{fig:morphologies_one_gap}, a dimer of two coalesced bridges
at the contacts in (b), three isolated pendular bridges in (c), and a
bridge dimer at the contacts coexisting with a pendular bridge across
the gap in (d). Throughout this work we assume that separate liquid
bodies can be exchange volume. Consequently, the relevant parameter in
the volume controlled case is the total volume of all liquid bodies.

Inspection of the liquid-vapor interface of the dimer and the trimer
shows that both interfaces are topologically equivalent to the surface
of a sphere perforated by three holes. Despite this similarity, the
overall shape of a trimer and dimer are qualitatively different. The
liquid of a trimer fills the central opening, or `throat', formed by
the three adjacent beads. In the presence of a dimer, however, the
central part of this throat is empty of liquid. Dimer morphologies can
thus be described by a pair of pendular bridges that have coalesced in
a small section of the contact lines.

\begin{figure}
  \centering
  \includegraphics[width=0.65\columnwidth]{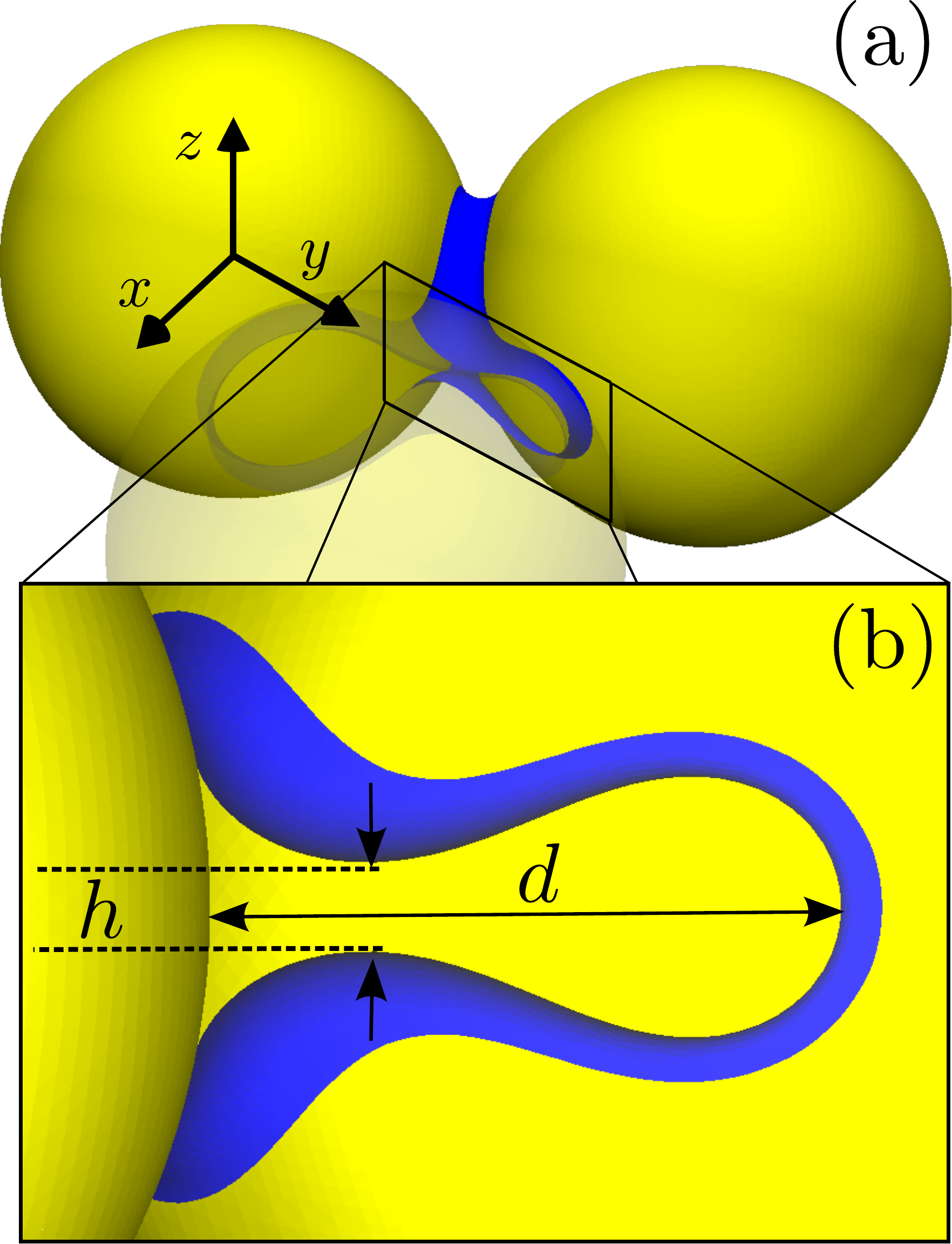}
  \caption{(a): View of a trimer morphology with liquid volume $V=0.2$
    and material contact angle $\theta_0=5^\circ$ wetting a throat of
    three beads with one finite gap $s_{12}>0$ and two contacts
    $s_{13}=s_{23}=0$. The foremost bead and half of the liquid
    interface are rendered transparent to enhance the cut along the
    symmetry plane. (b): Magnified view of a cut through a trimer
    along its symmetry plane, indicating the definition of the order
    parameter meniscus distance, $d$, and the distance $h$ of the two
    opposing menisci in the throat.}
  \label{fig:sketch_order_parameter}
\end{figure}

To distinguish the trimer from the dimer morphologies, and to detect
certain types of instabilities that occur during quasi-static changes
of the control parameter, we consider two suitable shape descriptors
as indicated in Fig.~\ref{fig:sketch_order_parameter}. Let us imagine
a cut through the liquid body by the symmetry plane $x=0$ orthogonal
to the plane $z=0$ passing through the centers of the spheres. The
first shape descriptor is now the distance $d$ between the
intersection of liquid interface and the $y$-axis, and the surface of
bead $3$, cf.~Fig.\ref{fig:sketch_order_parameter}. As a second shape
descriptor, we chose the smallest distance $h$ of two points on the
concave part of the upper and lower meniscus in the throat. This
minimum thickness of this `liquid lamella' may become ill-defined if
the meniscus in the throat lack a concave shape, unlike the example
shown in Fig.~\ref{fig:sketch_order_parameter}.

\begin{figure}
  \centering
  \includegraphics[width=0.85\columnwidth]{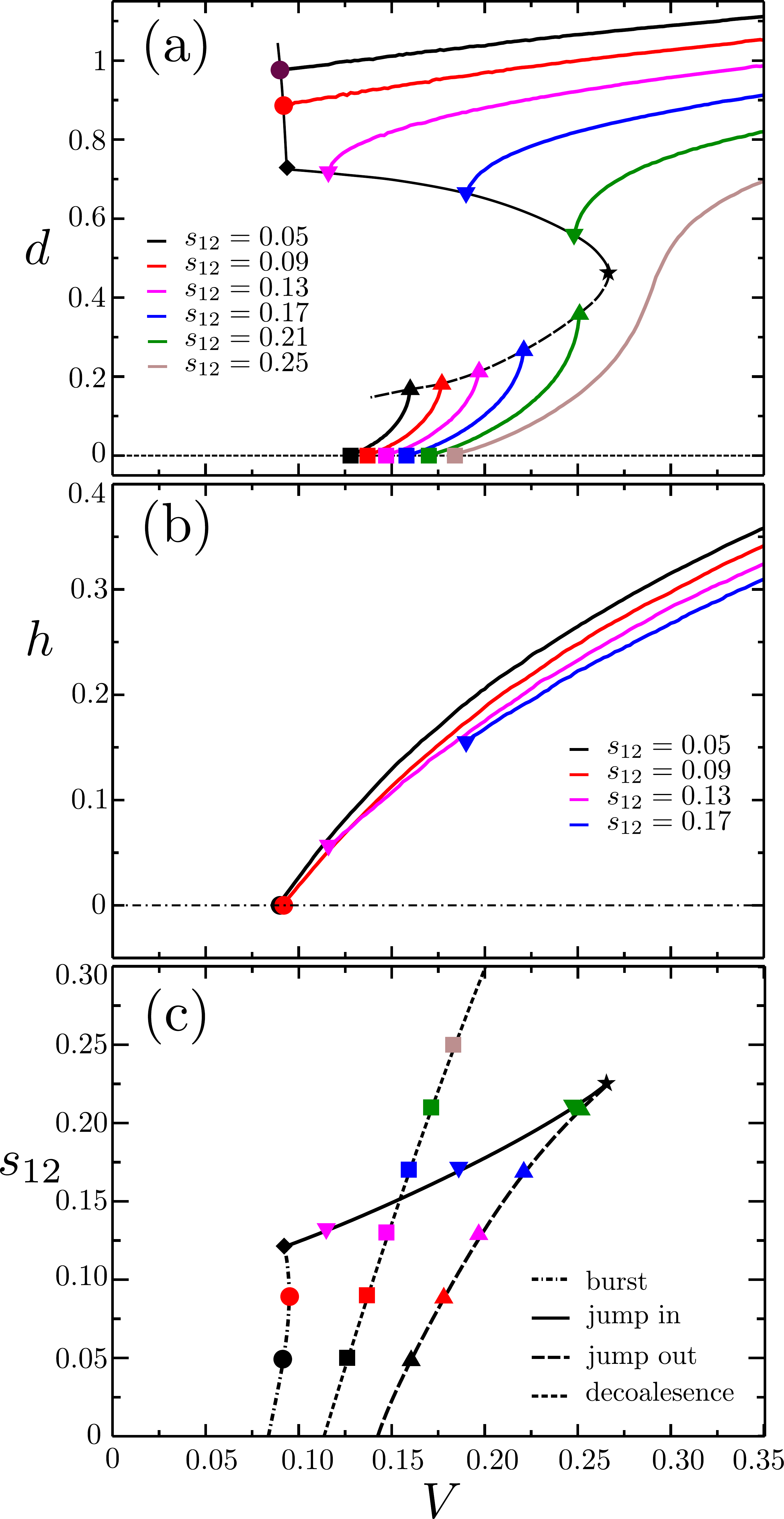}
    \caption{(Color online) (a): Distance $d$ between the outer
      meniscus at the gap and the surface of sphere $3$ as a function
      of the volume $V$ for different gap openings $s_{12}$. (b):
      Minimal distance $h$ of the two opposing menisci in the throat
      as function of $V$ for the same values of $s_{12}$. (c):
      Morphology diagram in terms of the $V$ and $s_{12}$ displaying
      the lines of instabilities as outlined in the main text. The
      bifurcation point and the kink are indicated by the star
      ($\star$) and diamond ($\filleddiamond$) symbol,
      respectively. For a definition of $d$ and $h$, cf.~also Fig.~
      \ref{fig:sketch_order_parameter}.}
  \label{fig:flip_flop_order_parameter}
\end{figure}

The plots in Fig.~\ref{fig:flip_flop_order_parameter}~(a) illustrate
the evolution of the meniscus distance $d$ as a function of the liquid
volume $V$, for a set of fixed gap opening $s_{12}$ in the range
between $0.05$ and $ 0.25$. At large a gap opening $s_{12}=0.25$, we
observe a continuous increase of $d$ for an increasing volume $V$, and
no discontinuous jumps. The distance $d$ follows the same curve during
a volume decrease and a clear transition between trimer at large $V$
and dimer at small $V$ cannot be found. Only at small volumes, the
trimer/dimer morphology eventually decays, caused by a de-coalescence
of the meniscus as $d\rightarrow 0$. This interfacial instability,
indicated by square symbols in
Fig.~\ref{fig:flip_flop_order_parameter}(a), will lead to two
separated pendular bridges as a final state.

At smaller gap openings $s_{12}=0.13,\,0.17,\,0.21$ the meniscus
distance $d$ shown in Fig.~\ref{fig:flip_flop_order_parameter}(a)
displays discontinuous jumps to larger or smaller values, depending on
whether the volume $V$ is decreased or increased, respectively. Owing
to these jumps, we find a range of control parameter $s_{12}$ and $V$
where two branches of local minima of the interfacial free energy
(\ref{eq:energy_functional_E}) can be found in our numerical energy
minimizations. This mechanical bistability between a dimer and a
trimer allows a clear distinction of the morphologies. The branch of
liquid conformations with the larger or smaller value of $d$ are
classified as trimers and dimers, respectively. The corresponding
interfacial instabilities limiting the range of mechanically stable
trimers and dimers will be termed the `snap--in' and the `pop--out'
instability in the remainder of this article. The latter two
instabilities are indicated by downward and upward triangles,
respectively, in Fig.~\ref{fig:flip_flop_order_parameter}.

For a gap opening $s_{12}$ of $0.13$ and smaller, the snap--in
instability occurs at a volume smaller than the smallest volume that
allows a stable dimer to be formed, cf.~the corresponding branches
shown in Fig.~\ref{fig:flip_flop_order_parameter}~(a). Hence, trimers
can directly break up into two separated pendular bridges during a
volume decrease even without passing through the dimer morphology. For
the examples shown in Fig.~\ref{fig:flip_flop_order_parameter}~(a),
the trimer branch terminates for a finite value $d>0$ at gap opening
$s_{12}$ of $0.05$ and $0.09$. In these cases the minimal liquid
thickness $h$ in the throat reaches zero before a discontinuous inward
jump of the outer meniscus in the gap or a de-coalescence can lead to
the decay of the trimer. A collision of the two opposing menisci in
the throat triggers a sudden opening of the liquid interfaces. The
circles in the plots shown in
Fig.~\ref{fig:flip_flop_order_parameter}~(a) and (b) indicate this
`burst' instability during a decreasing liquid volume. The burst
instability of the central liquid meniscus very likely leads to three
separated pendular bridges as the final state.

Figure~\ref{fig:flip_flop_order_parameter}~(c) illustrates the
stability boundaries of the trimer and dimer morphologies that
correspond to one of the four possible types of instability. With
systematic scans of parameters $s_{12}$ and $V$, we identified a
bifurcation point at $(V^\star, s_{12}^\star)=(0.26,0.22)$ as
indicated by the black star in the stability diagram
Fig.~\ref{fig:flip_flop_order_parameter}~(c). The tangents to the
lines indicating the snap--in and the pop--out instability become
parallel and terminate in a cusp. This type of bifurcation is
generically observed for two control parameters and one order
parameter, see e.g.~Ref.~\cite{Thom1989,Arnold2003}. The lines of the
snap--in instabilities and the burst emerge from a point
$(V^\filleddiamond, s_{12}^\filleddiamond)=(0.09, 0.125)$, where the
two stability lines form a kink. The latter point is indicated by a
black diamond in Fig.~\ref{fig:flip_flop_order_parameter}~(c). Note,
that mechanically stable dimers can exist in the `ideal' case
$s_{12}=0$ of all three beads in contact, provided that the volume
falls into the narrow range between $V\approx 0.115$ and $V\approx
0.15$.

It turns out that also a mechanically stable `chimera' morphology of a
dimer and an isolated pendular bridge across the finite gap is
possible. In contrast to the trimer, the latter morphology is always
metastable (a local energy minimum of the interfacial energy) and can
found only in a small region of control parameters. As for a single
bridge dimer, a volume decrease will likely induce a break-up of the
interface between the two coalesced bridges and the formation of three
pendular bridges. Alternative to de-coalescence, one may observe that
the pendular bridge located at the gap will transfer liquid into the
dimer and disappears.

An increase of the liquid volume of the trimer/dimer chimera state,
however, will lead to a coalesce of the dimer and the pendular bridge
at the gap. The resulting trimer will likely fill the throat
opening. Similar to the decay at decreasing volume an increase of the
gap opening at fixed volume could induce a rupture of the pendular
bridge at the gap. Here, we can speculate that the liquid will first
form a transient dimer that finally decays, after a pop--out of the
meniscus between the two coalesced bridges, into a trimer.

\subsection{Experiments}
\label{subsec:experiments}

\begin{figure}
  \includegraphics[width=0.95\columnwidth]{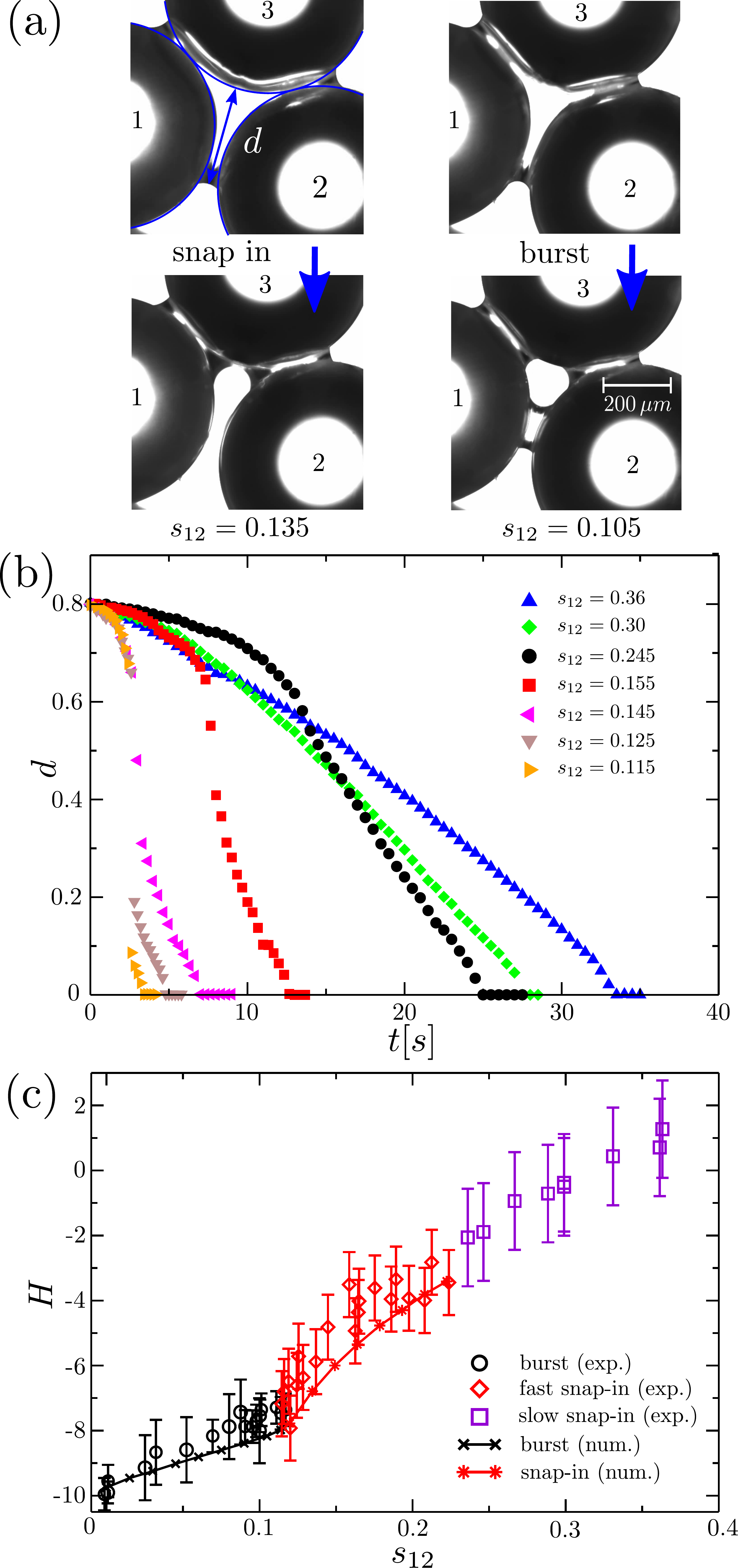}
  \caption{(Color online) Evaporation of pendular bridge trimer of
    three water bridges in assembly of three ruby beads with a radius
    of $R=(300\pm 5)\mu$m with a finite gap between bead 1 and
    2. (a): Optical micrographs of a pendular bridge trimer with a
    finite gap opening $s_{12}=0.115$ (top left) and $s_{12}=0.135$
    (top right) and the corresponding final states of the burst
    instability (left bottom) and snap--in instability (bottom
    right). (b): Distance $d$ of the outer meniscus in the gap from
    the surface of the opposing bead as a function of time $t$ elapsed
    after the meniscus has passed a distance $d=0.8$ for different gap
    opening $s_{12}$. (c): Estimated mean curvature $H$ of an
    evaporating trimer at the point of instability as a function of
    the gap opening $s_{12}$ in evaporation experiments and numerical
    energy minimizations.}
  \label{fig:two_decay_modes}
\end{figure}

Motivated by the results of our numerical energy minimizations in
Sec.~\ref{subsec:liquid_morphologies}, we investigated the decay of
the trimer morphology and the final states during a slow decrease of
liquid volume in an experimental realization. To this end, we fix
three spherical ruby beads of identical radius $R=(300\pm 5)\mu$m
(purchased from Saphirwerk Industrieprodukte AG and Sandoz Fils Sa,
both Switzerland) on a microscopy glass slide using play dough. Two
pairs of the beads are in mechanical contact ($s_{13}=s_{23}=0$) while
a small separation is intentionally left between the surfaces of the
third pair ($s_{12}>0$). Deionized water is employed as a volatile
wetting liquid with a receding contact angle of $\theta_{\rm
  r}\lesssim 10^\circ$ on the surface of the cleaned ruby beads,
similar to the material angle of $\theta_0=5^\circ$ in our numerical
energy minimizations in Sec.~\ref{subsec:liquid_morphologies}. In the
begining of the experiment a small water droplet is placed in the
throat formed by the three ruby beads. During evaporation the shape of
the meniscus is recorded by optical microscopy in top view, cf.~also
Fig.~\ref{fig:two_decay_modes}(a). Experiments are performed at
similar temperatures and relative humidities to ensure comparable
evaporation rates.

Examples of video frames of a trimer recorded during a typical
evaporation experiment are displayed in
Fig.~\ref{fig:two_decay_modes}(a) for two gap openings. In good
agreement with the numerical results of the previous section
\ref{subsec:liquid_morphologies}, we observe a burst instability for
$s_{12}=0.115$ (left column) and a discontinuous snap--in at
$s_{12}=0.145$ (right column). The distance of the meniscus in the gap
to the surface of the opposing bead is denoted as $d$, cf.~also
Fig.~\ref{fig:two_decay_modes}~(a) and
Sec.~\ref{subsec:liquid_morphologies}. According to our convention, we
rescaled the distance $d$ by the bead radius $R$.

Figure \ref{fig:two_decay_modes}~(b) shows the meniscus distance $d$
extracted from the video frames as a function of time $t$ for a series
of gap openings $s_{12}$. To compare the temporal evolution of $d$
during a snap--in for different values of the gap opening $s_{12}$, we
chose an individual offset on the time axis for each curve in
Fig.\ref{fig:two_decay_modes}~(b) such that the meniscus distance at
time $t=0$ attains the value $d=0.8$.

Irrespective of the gap opening $s_{12}$, the meniscus distance $d$ is
mono\-ton\-ous\-ly deceasing as time $t$ passes. The cross-over from a
sudden, discontinuous snap--in to a gradual and continuous decrease of
$d$ occurs between gap openings $s_{12}$ of $0.155$ and $0.245$. At
this point, we observe also a qualitative change in the form of the
functions shown in Fig.~\ref{fig:two_decay_modes}~(b), which display
an increasing s-shape of the curves as $s_{12}$ is decreased. The
transition agrees well with the predictions of our numerical energy
minimizations in Sec.\ref{subsec:liquid_morphologies}. An expected
bifurcation point at a gap opening $s_{12}^\star=0.22$ separates the
continuous from the discontinuous trimer decay. The corresponding
measured meniscus distance $d^\star$ at the expected bifurcation point
as well as the values of $d$ at the onset of the snap--in instability
for $s_{12}<s_{12}^\star$ agree well with the numerically obtained
results of Sec.~\ref{subsec:liquid_morphologies} shown in
Fig.~\ref{fig:flip_flop_order_parameter}.
 
The snap--in instability was not observed in our evaporation
experiments at gap openings $s_{12} \lesssim 0.115$. In the latter
cases, the trimer became unstable upon a volume reduction because the
central part of liquid-vapor undergoes a sudden burst instability. In
the majority of cases the final state attained after the snap--in
instability at $s_{12}>0.12$ is a pendular bridge dimer. In contrast
to the snap--in instability, burst instabilities occur for small gap
openings $s_{12}<0.12$, only, and lead to a final state of three
pendular bridges.

For a quantitative comparison of the trimer instabilities observed in
different experimental realizations and our numerical energy
minimizations, it is useful to consider the mean curvature $H$ of the
liquid-vapor interface at the onset of the instability leading to the
decay of the trimer. To this end we extracted the projected shape of
the menisci of the coalesced pendular bridges spanning the pair of
beads that are in contact. Because the outer parts of these interfaces
are hardly affected by the liquid in the throat connecting the
pendular bridges, the projected contours of these menisci are still
close to the shape of an isolated pendular bridge. Adopting the
toroidal approximation of pendular bridges for these menisci, we fit
the in--plane contour of the outer menisci by circular arcs. The mean
curvature $H$ of the interface is the arithmetic mean
$H=(\kappa_\parallel+\kappa_\perp)/2$ of the in-plane ($\parallel$)
and out-of-plane ($\perp$) curvatures $\kappa_\parallel=1/r_\parallel$
and $\kappa_\perp=1/r_\perp$, respectively. Here, $r_\parallel$ is the
radius of the fitted arc while $r_\perp$ denotes the distance of the
outer meniscus from the respective contact of the bead pair. This
geometrical analysis of the liquid interface is done only on the last
video frame that still shows a complete trimer, or before the outer
meniscus in the gap accelerated while approaching the snap--in
instability.

Figure \ref{fig:two_decay_modes}~(c) displays the values of the mean
curvature $H$ estimated in our experiments for various gap openings
$s_{12}$. In comparison to the experimental data, we plot in
Fig.~\ref{fig:two_decay_modes}~(c) the corresponding values of the mean
curvature obtained in our numerical energy minimizations. A transition
between the snap--in and the burst instability is clearly visible as a
kink in both the experimental and the numerical data. A small
systematic shift in the mean curvature between the experimental data
points and the numerical results is apparent in the plot of
Fig.~\ref{fig:two_decay_modes}~(c), but the numerical data fall into
the range of the experimental uncertainties. Our experimental data
show a cross-over between the burst and the snap--in instability for a
gap opening $s_{12}\approx 0.12$. This value coincides perfectly with
the numerically determined value where we assumed a material contact
angle of $\theta_0=5^\circ$.

\subsection{Capillary forces}
\label{subsec:capillary_forces}

After discussing the possible liquid states and their interfacial
instabilities in Sec.~\ref{subsec:liquid_morphologies} and
Sec.~\ref{subsec:experiments}, we will now turn to the capillary
forces on the beads in contact to the liquid. It is evident that the
total capillary force acting on a single bead in the presence of three
pendular bridges is a sum two central forces. But also for the
trimer/dimer morphology, the total forces $\bF_i$ acting on bead $i$
can be decomposed into a pairs of central forces, as illustrated in
Fig.~\ref{fig:force_decomposition_trimer}. Such a decomposition is
always possible for three spherical beads whose center points are not
co-linear. The proof is given in the Appendix
\ref{subsec:decomposition}. Due to the reflection symmetry of the bead
configuration, the magnitude of the pair forces $F_{ij}$ with $i,j \in
\{1,2,3\}$ are invariant upon interchanging bead $1$ by bead
$2$. Hence, the capillary forces satisfy $F_{13}=F_{23}$, and we need
to consider only two independent forces, the force $F_{12}$ acting
across the gap and the force $F_{13}$ acting at one of the two
contacts.

\subsubsection{Three pendular bridges}
\label{subsec:forces_three_bridges}

\begin{figure*}
  \includegraphics[width=0.8\textwidth]{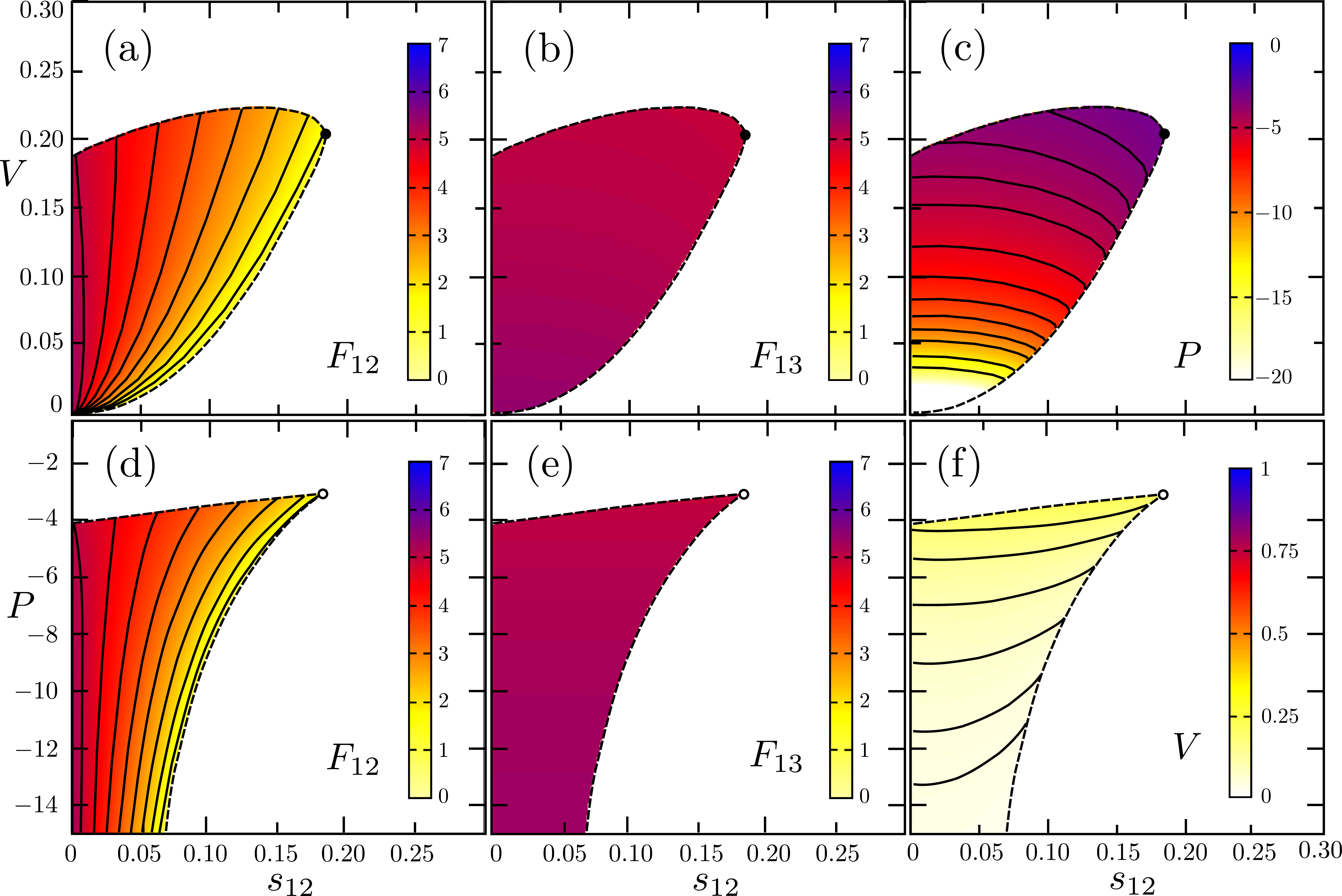}
  \caption{(Color online) Attractive capillary forces $F_{12}$ across
    the gap (a) and $F_{13}$ at the contact (b) in presence of three
    pendular bridges as a function of the gap opening $s_{12}$ and
    liquid volume $V$.  Capillary forces $F_{12}$ and $F_{13}$, and
    volume $V$ of three pendular bridges in the pressure controlled
    case are shown in panels (c), (d), and (f), respectively, as
    function of $s_{12}$ and $P$.}
  \label{fig:forces_bridges}
\end{figure*}

Figure \ref{fig:forces_bridges}~(a) and (b) illustrates the capillary
forces acting in a symmetric bead configuration with a finite gap
opening $s_{12}> 0$ and $s_{13}=s_{23}=0$ in the presence of three
pendular bridges with total volume $V$. As already mentioned in the
beginning of Sec.~\ref{subsec:liquid_morphologies}, we assume a mutual
exchange of liquid volume between the bridges. The Laplace pressure
$P$ of the bridges as a function of gap opening $s_{12}$ and total
volume $V$ is shown in Fig.~\ref{fig:forces_bridges}~(c).

The dashed lines in Fig.~\ref{fig:forces_bridges} indicate the region
$(s_{12},V)$ where configurations of three pendular bridges are
mechanically stable. Once the stability limit is reached during an
adiabatically slow change of $s_{12}$ and $V$, the bridge ensemble
will become unstable and decay into an interfacial morphology with a
lower interfacial free energy. For small total volumes $V$ below the
asterisk symbol in Fig.~\ref{fig:forces_bridges}, the ensemble of
three pendular bridges becomes unstable with respect to a mutual
exchange of volume. In the course of this instability, the pendular
bridge located at gap will be spontaneously `sucked up' by the two
bridges at the contacts. In the upper part of the stability limit
above the symbol, we observe that the contact lines of at least one
pair of bridges touch, followed by coalescence.

Inspection of Fig.~\ref{fig:forces_bridges}~(a) shows that the
attractive force $F_{12}$ acting across the gap at a fixed total
volume $V$ of the bridges varies strongly with the gap opening
$s_{12}$, dropping from a value $F_{12}\approx 5$ at $s_{12}=0$ to
$F_{12} \approx 0$ at the gap opening $s_{12}=s^\ast_{12}$ on the
stability limit for the given volume $V$. The values of $F_{12}$ at
the point $s_{12}=0$ and $s_{12}=s_{12}^\ast$ are rather insensitive
with respect to changes in $V$, cf.~Fig.~\ref{fig:forces_bridges}. As
expected, the force $F_{12}$ for a vanishing gap opening $s_{12}=0$
increases slightly with decreasing $V$, and only approaches the value
$2\pi\cos\theta_0\approx 6.23$ in the asymptotic limit $V\rightarrow
0$. The value of maximum gap opening $s^\ast_{12}$, however, depends
strongly on the total volume $V$.

In contrast to $F_{12}$, the attractive capillary force $F_{13}$ at
the bead contacts shown in Fig.~\ref{fig:forces_bridges}~(b) does not
display a significant variation with neither the gap opening $s_{12}$
nor the total volume $V$. As expected from the force $F_{12}$ at
$s_{12}=0$, we obtain values $F_{13}\approx 5.5$ with a slight
increase to $\approx 6.23$ in the asymptotic limit $V\rightarrow
0$. The Laplace pressure $P$ of three communicating pendular bridges
in Fig.~\ref{fig:forces_bridges}~(c) depends on both control
parameters, $s_{12}$ and $V$. Apparently, the dependence of $P$ on $V$
is more pronounced than the dependence on $s_{12}$.

\begin{figure*}
  \includegraphics[width=0.8\textwidth]{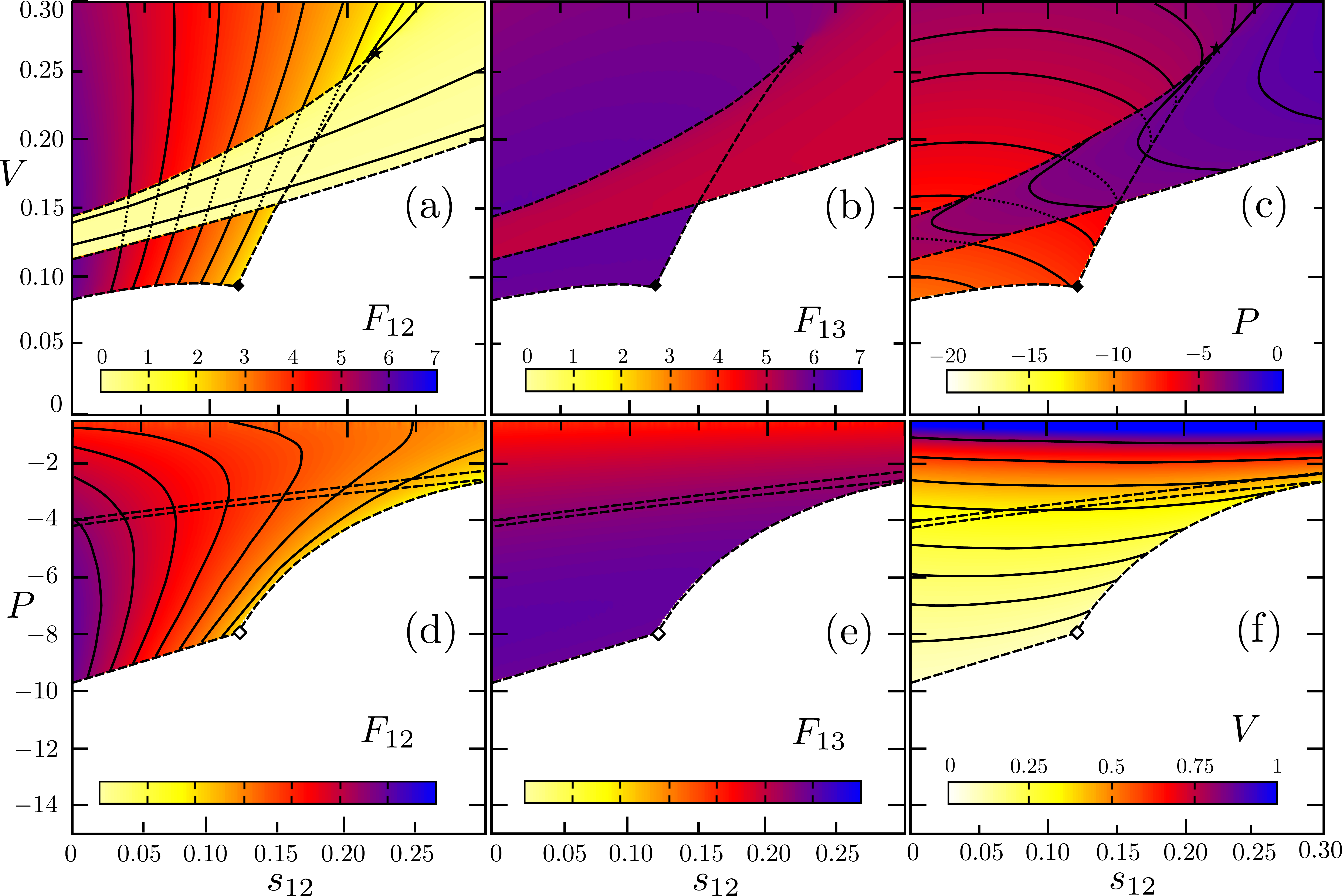}
  \caption{(Color online) Attractive capillary forces $F_{12}$ across
    the gap (a), $F_{13}$ at the contact (b), and Laplace pressure $P$
    of trimer/dimer (c) as a function of the gap opening $s_{12}$ and
    liquid volume $V$. In regions of trimer/dimer bistability, the
    forces of the morphology with the smaller meniscus distance $d$ is
    displayed, cf.~also the diagram in
    Fig.~\ref{fig:flip_flop_order_parameter}. Capillary forces
    $F_{12}$ and $F_{13}$, and volume $V$ of a trimer/dimer morphology
    in the pressure controlled case are shown in panels (d), (e), and
    (f), respectively, as function of $s_{12}$ and $P$ .}
  \label{fig:forces_trimer}
\end{figure*}

In many instances it is appropriate to assume that the clusters in the
wet granular assembly exchange liquid with neighboring structures
\cite{Scheel2008a,Scheel2008b,Mani2015}. Here, one may regard the
average Laplace pressure of the neighboring structures as the relevant
control parameter. In the latter case, we need to consider local
minima of the Grand interfacial energy ${\cal G}$ rather than local
minima of the interfacial energy ${\cal E}$.

Capillary forces $F_{12}$ and $F_{13}$ of three pendular bridges in
the pressure controlled case are shown in
Fig.~\ref{fig:forces_bridges}~(d)and (e) as a function of gap opening
$s_{12}$ and Laplace pressure $P$. We find the same cross-over between
the decay modes as in the volume controlled case discussed above.  For
high Laplace pressures $P$ (with negative sign, but with low
magnitude), the pendular bridges are unstable with respect to
coalescence while for small values of $P$ (i.e. with negative sign,
and large magnitude), the bridge at the gap becomes unstable with
respect to a volume exchange with the reservoir. The corresponding
stability limit in Fig.~\ref{fig:forces_bridges} is shown as a dashed
line where the symbol indicates the cross-over between the two modes
of instability.

Similar to the volume controlled case, the force $F_{12}$ acting
across the gap decays with increasing gap opening $s_{12}$, cf.~Fig
\ref{fig:forces_bridges}~(d). At a fixed gap opening $s_{12}$, the
capillary forces become stronger with a decreasing Laplace pressure
$P$ (i.e. with negative sign and increasing magnitude), reaching the
asymptotic value $2\pi\cos\theta\approx 6.23$ for $s_{12}=0$ only in
the limit $P\rightarrow -\infty$. Close to the segment of the
stability boundary related to bridge coalescence, we find
$F_{12}\approx 5$, similar to the values in the volume controlled case
at large $V$. Figure \ref{fig:forces_bridges}~(e) displays the
capillary force at the bead contacts, $F_{13}$, which is insensitive
with respect to $P$, and with a similar values in the range
$F_{13}\approx 5$ close to the coalescence line. The total volume $V$
of the three bridges is shown the last panel of
Fig.~\ref{fig:forces_bridges}~(f).

\subsubsection{Trimer/dimer morphology}
\label{subsec:forces_trimer_dimer}

In the following we will discuss the capillary forces of the
trimer/dimer morphology in the volume and pressure controlled
cases. Figure~\ref{fig:forces_trimer}~(a) and (b) display the
magnitude of the attractive capillary forces $F_{12}$ and $F_{23}$ In
regions of the control parameter gap opening $s_{12}$ and liquid
volume $V$ related to a morphological bistability, i.e.~where both the
dimer and the trimer are locally stable shapes, the map displays the
forces corresponding to the liquid conformation with the larger value
of the order parameter $d$, which is the trimer.

In contrast to a trimer a dimer exerts a very small attractive force
$F_{12}$ across the gap as expected from the liquid distribution
between the beads. The attractive force $F_{12}$ in presence of a
trimer depends strongly on the gap opening $s_{12}$ but only weakly on
the volume $V$, while its magnitude is much larger than that of a
dimer. Only small variations of the capillary force $F_{13}$ at the
two contacts is observed with respect to variations of the gap opening
$s_{12}$. The attractive force $F_{13}$ with a value around $6$ hardly
depends on the volume $V$ and is not far from the asymptotic value
$2\pi\cos\theta_0\approx 6.23$ for pendular bridges between two
spherical beads for $s_{12}=0$ in the limit $V\rightarrow 0$. Similar
to the trimer, the dimer exerts rather constant forces between the
spherical beads in contact which is approximately $F_{13}\approx 5$,
i.e.~less than the capillary force in presence of a trimer with the
identical volume.

Figure~\ref{fig:forces_trimer}~(c) displays the Laplace pressure $P$
of the dimer/trimer morphology. The Laplace pressure mainly depends on
the liquid volume $V$ and only weakly on the gap opening
$s_{12}$. Note, that dimers display a larger Laplace pressure
(i.e.~negative with a smaller magnitude) as compared to a trimer at
the same volume. A transfer of liquid from the outer menisci of the
dimer into the central region of the throat is accompanied by a
decrease of volume in the outer menisci. The redistribution of liquid
volume leads to a decrease of the mean curvature and, hence, to a
decrease of the Laplace pressure. Note, that the Laplace pressure of a
trimer is typically negative but may become positive for large
volumes.

Figure \ref{fig:forces_trimer}~(d) and (e) display the capillary force
$F_{12}$ at the gap and $F_{13}$ at the contacts for liquid
morphologies in the plane spanned by the gap opening $s_{12}$ and
Laplace pressure $P$. It is apparent that the region in $(s_{12},
P)$-plane where the dimer/trimer morphology exists as a local minimum
of the Grand interfacial free energy displays a shape similar to the
region in the volume controlled case. As expected, the `burst' and the
snap--in instability modes form a kink in the stability limit of the
trimer at the point $(s_{12}^\diamond, P^\diamond)=(0.12,-8.0)$. The
two almost parallel dashed lines in the upper region of the color
plots indicate the narrow region where dimers can exist as metastable
configurations in the pressure controlled case. As in the volume
controlled case, trimers can decay either by the snap--in of the
liquid menisci at the sides or by a burst instability leading to an
opening of the throat. The pressure at the latter instability linearly
increases with the gap opening $s_{12}$, and the curve $P(s_{12})$
terminates together with the lines of the snap--in and the pop--out
instability in a point $(s_{12}^\diamond,P^\diamond)=(0.12,-8.5)$. At
this point, the stability boundary of the trimer exhibits a kink. The
stability boundaries corresponding to the snap--in and pop--out
instabilities join smoothly in a cusp bifurcation point at
$(s_{12}^\ast,P^\ast)=(0.4,-2.3)$ which lies outside the range shown
in the Fig.~\ref{fig:forces_trimer}(d) to (f).

The capillary force $F_{12}$ acting across the gap in the presence of
a trimer displays a strong dependence on the gap opening
$s_{12}$. However, the magnitude of $F_{12}$ varies only weakly with
the Laplace pressure $P$. A stronger dependence is observed only above
a value of $P\approx -4.5$, at a point where the trimer itself would
have already fused with other neighboring liquid structures (bridges or
trimer) inside the random assembly of spherical beads
\cite{Scheel2008a,Scheel2008b}. The capillary force $F_{12}$ at the
gap created by the trimer is approximately by a factor $1.1$ larger
than the capillary force of a single bridge at a prescribed Laplace
pressure of $P=-4.5$ at the same distance $s_{12}$.

Figure \ref{fig:forces_trimer}~(e) reveals that the capillary force
$F_{13}$ at the contacts depends on the magnitude of the Laplace
pressure $P$, while $F_{13}$ is virtually constant with respect to the
gap opening $s_{12}$. Also in this case, the capillary force is
approximately by a factor $\approx 1.1$ larger than the force for a
single pendular bridge at the contact held at the same value of
$P=-4.5$. The liquid volume $V$ of a trimer in the pressure controlled
case depends to a larger degree on the magnitude of $P$ and only
weakly on the gap opening $s_{12}$, cf.~the color plot in
Fig.~\ref{fig:forces_trimer}~(f). An analogous statement applies to an
ensemble of three isolated pendular bridges that are allowed to
exchange liquid.

\begin{figure*}
  \includegraphics[width=0.65\textwidth]{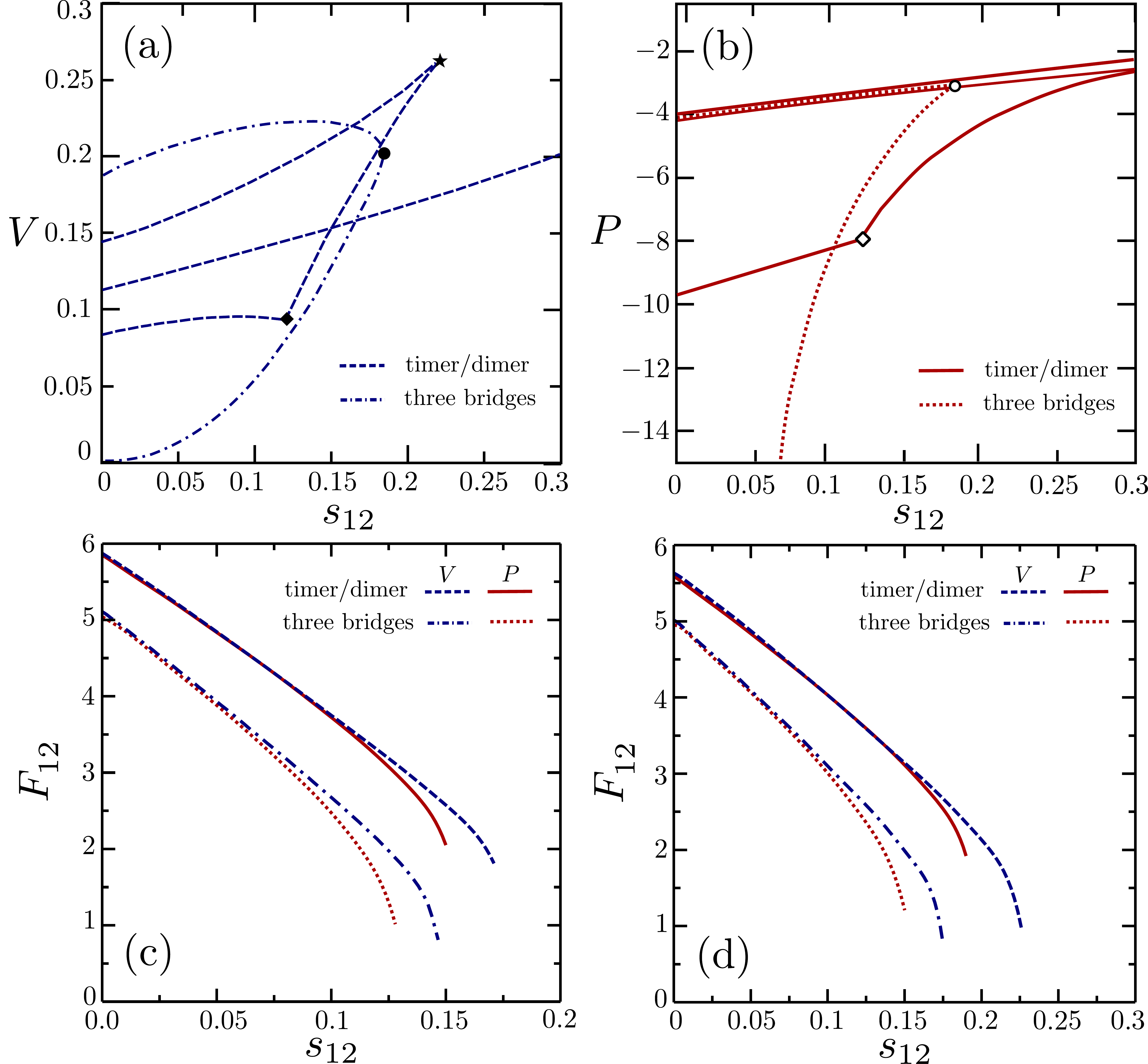}
  \caption{(Color online) Comparison of the stability boundaries of a
    trimer/dimer morphology and of three pendular bridges in terms of
    the gap opening $s_{12}$ and (a) the total liquid volume $V$ in
    the volume controlled case or (b) the Laplace pressure $P$ in the
    pressure controlled case. (c): capillary force $F_{12}$ at the gap
    as a function of gap opening $s_{12}$. Shown are data for three
    pendular bridges at a fixed total volume $V=0.116$ (dashed-dotted
    lines) and for a fixed Laplace pressure $P=-6$ (dotted line) in
    comparison to Data for the trimer morphology at a fixed volume
    $V=0.190$ (dashed line) and for a fixed Laplace pressure $P=-6$
    (solid line). (d): Data for three pendular bridges of fixed total
    volume $V=0.1644$ (dashed-dotted line) and for a fixed Laplace
    pressure $P=-4.5$ (dotted line) in comparison to data for a trimer
    at fixed volume $V=0.264$ (dashed line) and for a fixed Laplace
    pressure $P=-4.5$ (solid line).}
  \label{fig:comparison}
\end{figure*}

Simulations of the mechanics and dynamics of large wet granular
assemblies require efficient models for capillary cohesion. At low
liquid saturations, in the regime of pendular bridges, the cohesive
force can be treated as a superposition of two--body forces. For 
saturations in the funicular regime, a simple mapping of the 
capillary cohesion forces onto an equivalent ensemble of pendular 
bridges would be desirable. In the most simple approach, one would 
assume that the magnitude of capillary forces caused by the trimer/dimer
are similar to those in the presence of three pendular bridges for the 
same volume or Laplace pressure.

Figure \ref{fig:comparison}~(a) and (b) provides a quantitative
comparison of the stability boundaries of three pendular bridges to
the ones of the trimer/dimer morphology in the volume controlled and
the pressure controlled cases, respectively. Figure
\ref{fig:comparison} (c) and (d) display the capillary force at the
gap for three pendular bridges and the trimer/dimer morphology as a
function of the gap opening. As we are plotting the capillary force
$F_{12}$ and $F_{13}$ in the volume controlled and the Laplace
pressure controlled case in the same plot as a function, we chose
specific values of volume $V$ and Laplace pressure $P$.  For these
particular values the morphologies and, hence, the capillary forces
are identical at zero gap opening $s_{12}=0$.

Figure \ref{fig:comparison} (c) and (d) clearly demonstrate that a
trimer induces a higher attractive force across the gap. In addition,
the snap--in transition from trimer to dimer occurs at a larger gap
opening as compared to the point where the bridge at the gap decays in
a ensemble of three communicating bridges.

\section{Conclusion and outlook}
\label{sec:outlook}

In this article we have explored the morphology and capillary forces
of a pendular bridge trimer representing the most fundamental liquid
cluster in disordered assemblies of wet spherical beads. We have
focused our study on local triangular bead configurations with two
contacts and one finite gap. Numerical minimizations of the
interfacial energy reveal a shape bistability between a dimer and a
trimer of pendular bridges. Such a capillary hysteresis between the
trimer and the dimer shapes is present both in the volume and in the
Laplace pressure controlled cases. As the gap opening is changed, we
find a cross--over between two different interfacial instabilities
that can occur during a quasi-static volume reduction. At small gap
openings the two opposing menisci in the triangular throat touch and
break up in a sudden burst. For intermediate gap openings, however,
the timer decays by a sudden, discontinuous snap--in of the meniscus
in the gap once the meniscus is close to the narrowest point. A
continuous motion of the meniscus is observed in wide gaps. Systematic
evaporation experiments in the geometry of three beads with a single
finite gap opening quantitatively reproduce the transition from a
snap--in and a burst instability of a trimer as the gap opening is
decreased.

In respect to the modeling of the attractive capillary forces, it is
well justified to replace the bridge dimer by two separate pendular
bridges which are held at the same Laplace pressure as the
dimer. Bridge trimers, however, induce attractive capillary forces
across the gap that are slightly larger than the force of a pendular
bridge at the gap for the same Laplace pressure. The rupture distance
of the pendular bridge at the gap is systematically larger than the
gap opening where the snap--in transition from trimer to dimer
occurs. Hence, trimers of pendular bridges not only yields to an
increased cohesion between the beads but also the range over which the
increased capillary force act is enlarged. Both effects lead to a
enhanced dissipation of work during a slow deformation of a granular
assembly. These two observation may already be sufficient to explain
the shallow maximum of the strength of wet granulates reported in
experiments \cite{Scheel2008a,Scheel2008b}.

\acknowledgments{The authors would like to thank R.~Mani, D.~Kadau,
and H.-J. Herrmann for fruitful discussions. This work has been funded
by the German Science Foundation (DFG) within the SPP 1486 `PiKo'
under Grant No.~HE 2016/14-2.}

\section*{Appendix}

\subsection{Equivalence of capillary forces}
\label{subsec:ensemble_capillary_forces}

Respecting the sign convention of the Laplace pressure, we find the
identity
\begin{equation}
  P=\tilde P(\bar\br,V)=\partial_V E(\bar\br,V)~.
  \label{eq:Laplace_equation_derivative}
\end{equation}
Starting from the definition (\ref{eq:energy_functional_G}) of the
Grand interfacial free energy $\cal G$ and differentiation with
respect to the coordinates $\bar\br$ and using
eqn.~(\ref{eq:Laplace_equation_derivative}) we obtain the following
identity:
\begin{equation}
  \grad_{{\br}_i} G\big(\bar\br,P \big) 
  = \grad_{{\br}_i} E\big(\bar\br,\tilde V(\bar\br,P)\big)
  -P\,\grad_{{\br}_i}\tilde V\big(\bar\br,P\big)~.
  \label{eq:force_G}
\end{equation}
By the chain rule of differentiation, we can write the first term in
eqn.~(\ref{eq:force_G}) as
\begin{eqnarray}
  \lefteqn{\grad_{{\br}_i}E\big(\bar{\br},\tilde V(\bar{\br},P)\big)
    =\grad_{{\br}_i} E\big(\bar\br,V)|_{V=\tilde
      V(\bar\br,P)}}\nonumber \\ &+&\partial_V E(\bar\br,V)|_{V=\tilde
    V(\bar\br,P)} \grad_{{\br}_i}\tilde V\big(\bar{\br},P \big)~,
  \label{eq:total_force_energy_G}
\end{eqnarray}
and using the expression eqn.~(\ref{eq:Laplace_equation_derivative})
for the Laplace pressure, we finally obtain
\begin{equation}
  \grad_{{\br}_i}G\big(\bar\br, P \big)=
  \grad_{{\br}_i}E\big(\bar\br,V \big)|_{V=\tilde V(\bar\br, P)}~,
  \label{eq:total_force_energy_equiv_app}
\end{equation}
which proofs the equivalence of capillary forces in the volume and
pressure ensembles.

\subsection{Decomposition of forces}
\label{subsec:decomposition}

First, we will consider the case of three identical beads whose
centers are not co-linear, i.e.~$\br_{12} \times \br_{13}\neq {\mathbf
  0}$ where $\br_{ij} \equiv \br_i-\br_j$ are the relative positions
of the bead centers $\br_i$, $i\in\{1,2,3\}$. The co-linear case will
be discussed separately.

Employing the unit vector $\bn_{ij}$ pointing from the center of bead
$i$ to the center of bead $j$, and the unit vector $\bn^\perp$ given
by
\begin{equation}
  \bn^\perp=\frac{\br_{12}\times \br_{13}}{|\br_{12}\times \br_{13}|}~,
\end{equation}
we can decompose the forces $\bF_i$ onto bead $i$ into a sum of forces
\begin{equation}
  \sum_{i=1 \atop j\neq i}^3\,F_{ij}\bn_{ij}+F_i^\perp\,\bn^\perp=
      {\mathbf 0}
  \label{eq:force_sum_three_beads}
\end{equation} 
with $j\in\{1,2,3\}$ and $j\neq i$. To proof the conjecture that the
three body force can be written as sum over central force pairs, we
now have to show that $F_{ij}=-F_{ji}$ as well as $F_i^\perp=0$ holds
for all $i,j\in\{1,2,3\}$. We start our proof with the observation
that the total torque on the beads vanishes:
\begin{equation}
  \sum_{i=1}^3\;\left(\br_i\times\bF_i+\bT_i\right)={\mathbf 0}~.
  \label{eq:torque_sum_three_beads}
\end{equation}
Under the assumption that $\bT_i={\mathbf 0}$ for $i\in\{1,2,3\}$, we
are left with the identity
\begin{equation}
  \sum_{i=1}^3\;\mathbf{M}\cdot
  \left[\,\bF_i\times\left(\br_i-\br_0\right)\right]=0 
  \label{eq:torque_axis}
\end{equation}
for an arbitrary rotation axis $|\mathbf{M}|=1$ and center of
rotation, $\br_0$, because all forces $\bF_i$ with respect to the
centers of the beads must sum to zero. Without restricting generality,
we chose the center $\br_0 \equiv \br_1$ and a direction $\mathbf{M}
\equiv \bn^\perp$ of the axis. Equation (\ref{eq:torque_axis}) can be
now rewritten as
\begin{eqnarray}
  \bn^\perp\cdot &&\left[(F_{23}\,\bn_{23}+
    F_2^\perp\,\bn^\perp)\times
    \br_{13}\right.\nonumber\\ &&\left.(F_{32}\,\bn_{32}+F_3^\perp\,\bn^\perp)
    \times \br_{12}\right]=0~.
  \label{eq:torque_axis_b}
\end{eqnarray} 
By the definition of the unit vectors $\bn^\perp$ and
$\bn_{23}=-\bn_{32}$, we have $\bn^\perp\cdot \bn_{ij}=0$ and
\begin{equation}
  \bn_{23}\times \br_{13}=-\bn_{32}\times \br_{12}= \frac{2\,A_\Delta
  }{|\br_{23}|}\;\bn^\perp~,
  \label{eq:levers}
\end{equation} 
where $A_\Delta$ is the area of the triangle defined by the three bead
centers. With eqn.~(\ref{eq:levers}) and
eqn.~(\ref{eq:torque_axis_b}), we arrive at $F_{23}=-F_{32}$. The
choice of bead indices was arbitrary which implies that also
$F_{12}=-F_{21}$ and $F_{13}=-F_{31}$ must hold.

In order to show $F_1^\perp=F_2^\perp=F_3^\perp=0$, we chose the
rotation axis in the plane of the three bead centers, passing through
the center of bead $1$. The direction of the rotation axis is given by
\begin{equation}
  {\mathbf M}\equiv \frac{\br_{21}+\br_{31}}{|\br_{21}+\br_{31}|}
\end{equation}   
and $\br_0 \equiv \br_1$. With this choice, the torque acting on the
beads has to satisfy
\begin{equation}
  \frac{(\br_{21}+\br_{31})\times \bn^\perp (F_2^\perp-F_3^\perp)}
       {|\br_{21}+\br_{31}|}=0
\end{equation} 
which gives $F_2^\perp=F_3^\perp$. 

Following the same line of arguments for a rotation axis passing
through the center of bead $2$ and the point
$\br_2+(\br_{12}+\br_{32})/2$, we obtain $F_1^\perp=F_2^\perp$. Since
the sum of over all forces $\bF_{ij}$ in the plane of the three bead
centers vanishes, also all normal forces must sum to zero, leaving as
the only possibility $F_1^\perp=F_2^\perp=F_3^\perp=0$.

In the case of three co-linear bead centers, a unique decomposition
into three central force pairs must not necessarily be possible. A
simple counterexample are forces $\bF_1=\bF_3$ and $\bF_2=-2\,\bF_1$,
all orthogonal to the line, distances $\br_{12}=\br_{32}$ where bead
$2$ is located halfway in between beads $1$ and $3$. The singular
nature of this case can be easily seen when approaching the co-linear
configuration from a configuration with a slight bend in the plane of
the forces. If there is no spontaneous symmetry breaking of the liquid
shape, all capillary forces must act parallel to line passing through
the bead centers which, again, allows a unique decomposition into
force pairs.



\end{document}